# The Cathedral and the Starship:

# Learning from the Middle Ages for Future Long-Duration Projects

Andreas M. Hein

Andreas.hein@i4is.org, Initiative for Interstellar Studies (i4is), 27/29 South Lambeth Road, London, SW8 1SZ, United Kingdom

**Abstract**

A popular analogue used in the space domain is that of historical building projects, notably cathedrals that took decades and in some cases centuries to complete. Cathedrals are often taken as archetypes for long-term projects. In this article, I will explore the cathedral from the point of view of project management and systems architecting and draw implications for long-term projects in the space domain, notably developing a starship. I will show that the popular image of a cathedral as a continuous long-term project is in contradiction to the current state of research. More specifically, I will show that for the following propositions: The cathedrals were built based on an initial detailed master plan; Building was a continuous process that adhered to the master plan; Investments were continuously provided for the building process. Although initial plans might have existed, the construction process took often place in multiple campaigns, sometimes separated by decades. Such interruptions made knowledge-preservation very challenging. The reason for the long stretches of inactivity was mostly due to a lack of funding. Hence, the availability of funding coincided with construction activity. These findings paint a much more relevant picture of cathedral building for long-duration projects today: How can a project be completed despite a range of uncertainties regarding loss in skills, shortage in funding, and interruptions? It is concluded that long-term projects such as an interstellar exploration program can take inspiration from cathedrals by developing a modular architecture, allowing for extensibility and flexibility, thinking about value delivery at an early point, and establishing mechanisms and an organization for stable funding.

## 1. Introduction

Long-term space exploration programs such as interstellar exploration are frequently compared to the construction of monumental buildings such as cathedrals that took decades to centuries to complete [1,2]. For example, former NASA Administrator Michael Griffin declares that we owe cathedral-builders "the ability to have a constancy of purpose across years and decades." [2]

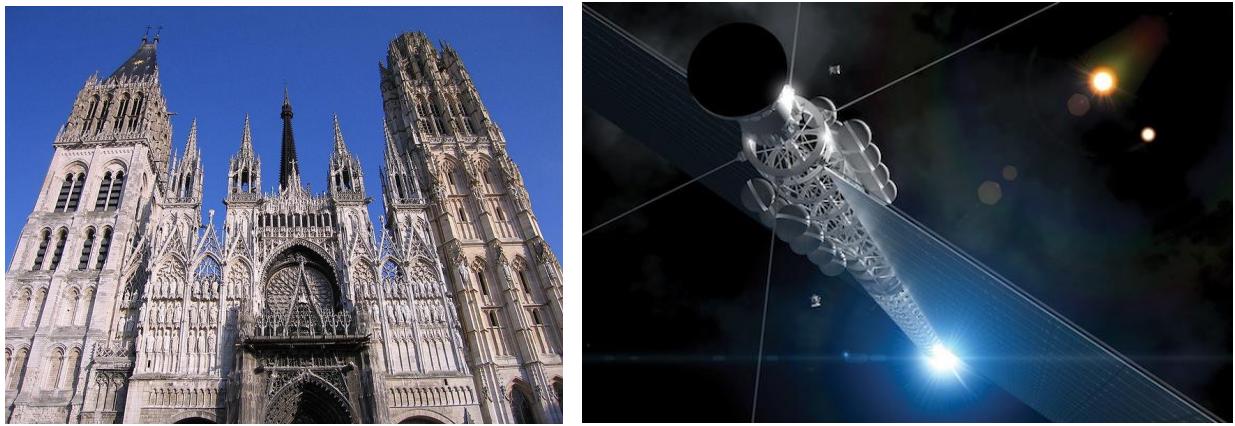

*Figure 1: Rouen Cathedral and a starship (Credit: Wikipedia, Adrian Mann)*

Cathedral builders "learned how to organize large projects, a key to modern society. And, probably most important of all, the cathedrals had to be, for decades at a time, a focus of civic accomplishment and energy." He continues to make the link between cathedrals and space programs by claiming that "the products of our space program are today's cathedrals." Krafft Ehricke, one of the space pioneers who developed the Extraterrestrial Imperative notes that "Like the giant cathedrals of the Middle Ages, Selenopolis will be the work of many generations" [3]. In a similar spirit, a radio show StarTalk by Neil deGrasse Tyson had the title "The International Space Station – A Modern Age Cathedral".

Putting cathedrals and space projects side by side, as in Figure 1, what are the implications? In these exemplary quotations, the cathedral symbolizes at least three things:

- A collective long-term achievement;
- A feat in long-term project management and engineering;
- A monument lasting for future generations.

Obviously, the analogy only works if these assertions regarding cathedrals are actually true. Among the three assertions, the third seems to be rather uncontroversial. Few will deny the cultural significance of cathedrals as monuments. The reaction to the Notre-Dame de Paris fire in 2019 on a global level seems to confirm that the cathedral has an enormous symbolic value, independently of individual religious beliefs. Hence, those who built the cathedral undeniably created a building which is a lasting monument many generations after construction finished.

It is important to define the key concepts that are used in the following. First, a cathedral is a church which contains the seat of the bishop. The bishop is a member of the Christian clergy. In the cases I consider in the following of the clergy of the Roman Catholic Church. The bishop's authority stretches over a certain geographic domain, the diocese. The bishop is accompanied by a group of clerics that he needs to consult in all important matters. This group is called the cathedral chapter. The cathedral chapter is important, as it was often responsible for the long-term management of cathedral-building and ensured its continuity via its underlying endowment fund, called "fabric". The bishop was often involved in the initiation of construction but at the later stages of the Middle Ages, the cathedral chapter was mostly responsible for the oversight of cathedral construction.

Cathedrals were far away from being a pure monument without utility value. "Thus, far from simply being a place of worship, the medieval cathedral was a multi-purpose structure, with some of the characteristics of the modern commercial mall, union hall, courthouse and amusement park" [6, p.455] "The role of the medieval cathedral was multi-faceted, with religious observance and ritual mixed with civil and commercial interests." [6, p.456] Thus, considering cathedrals for their symbolic value alone would fall short. This is even more important when it comes to the value of the cathedral in relation to its cost.

In the following, I would like to focus on the point of view of project management, project finance, and systems architecting. There is a rich literature on building cathedrals. Examples are [5–7]. Also, the management of these projects was also subject to several publications by Turnbull [8–11], who asserts that gothic cathedrals were laboratories, where construction was a process of experimentation. Chiu [12] looks at cathedrals from the perspective of a history of project management. Financing of cathedral building has also been subject to several works, notably Vroom's hallmark work, which presents several data sets of the financing history of several cathedrals [13]. Further publications have provided more or less quantitative accounts of cathedral financing [14,15].

Such an analysis would not only confirm or not confirm the appropriateness of using the cathedral as an analogy, but could also provide valuable insights into the drawing possible conclusions for current and future long-term projects, for example in spaceflight.

I first clarify some key concepts for understanding cathedral building. Then I conduct survey of the literature, combined with a statistical analysis of how long cathedral-building actually took, using a sample of French and English cathedrals. From these insights, I will try to develop implications that can be generalized and potentially applied to future long-duration projects in the space domain.

## 2. Materials and methods

For the cathedral to be a long-term achievement in project management, including financing, requires the existence of a long-term project. In other words, at least some form of continuity must exist. According to the dictionary definition, a project is "an individual or collaborative enterprise that is carefully planned to achieve a particular aim" (Oxford Dictionary). I will interpret "carefully planned" in the following way: For a cathedral to be a long-term project, at least some plan must exist. For the aim of the project to be reasonable, adequate means need to be provided as well. This leads me to the following hypotheses, I would like to test:

- The cathedrals were built based on an initial detailed master plan;
- Construction was a continuous process that adhered to the master plan;
- Investments were continuously provided for the building process.

I use the existing literature and elemental statistics to answer the research questions pertaining to the cathedral. However, I will frame or express the questions and answers in the language of project management and systems engineering. I argue that this framing yields conclusions that can be sufficiently generalized to apply to current and future long-term, large-scale engineering projects. I am aware that such a reframing needs to be done with care, as such a translation from one context to another inevitably introduces losses and bias. Whenever adequate, I will make this process as transparent as possible. Did cathedral-building proceed according to an initial detailed master plan?

## 3. Were cathedrals built according to a master plan?

What makes this question difficult to answer is that no coherent set of plans for a cathedral has been conserved. Hence, it is not possible to compare an initial plan with the actually realized cathedral. However, the question can be answered indirectly. If the cathedral had a master plan, it would most likely have strived towards coherence, harmony, and symmetry. Hence, the occurrence of multiple architectural styles and asymmetry would hint at a divergence from the initial plans if any existed. Indeed, this is what can be found in most cathedrals, notably in cathedrals that have been built over centuries such as the Rouen Cathedral. For example, the towers of Chartres Cathedral were built during the 12th and 16th century respectively, and their design differs considerably. But even the nave and choir, the main part of the cathedral that was built during a relatively short period of decades shows considerable differences in its construction and style of, e.g. the flying buttresses [18, p.274]. These observations hint at continuous modifications of the design of new elements of the cathedral. Such modifications might be correcting errors of a previous architect [18, p.276], changes in style, and improving skills of architects. Furthermore, modifications were necessary when funding ran out and the original plan could no longer be executed [5,

p.138]. As Scott [5] and Turnbull [11] remark, at least before the 13th century, detailed construction drawings with the right proportions and scale did not exist. Although models might have existed for discussions between master masons and the clergy, they were likely small and did not exhibit much detail. These results suggest that the design of cathedrals underwent continuous modifications of newly built elements. While a high-level design might have existed, prescribing the location of significant elements of the cathedral (e.g. towers), the detailed design of the actual elements of the cathedral could not have been prescribed in advance.

## 4. Was building a continuous process?

To start with, I will show via a simple statistical analysis that cathedrals were indeed built over centuries and then analyze how far the building process was continuous or not. I select a sample of 21 cathedrals built in the Middle Ages (Construction started before 1600), shown in Table 1.

*Table 1: Sample cathedrals for which construction started before 1600*

|  | **Begin of construction** | **End of construction** | **Total duration construction** |
| --- | --- | --- | --- |
| Rouen | 1145 | 1880 | 735 |
| Notre Dame | 1163 | 1345 | 182 |
| Strasbourg | 1176 | 1439 | 263 |
| Cologne | 1248 | 1880 | 632 |
| Rodez | 1276 | 1531 | 255 |
| Reims | 1211 | 1345 | 134 |
| Quimper | 1239 | 1515 | 276 |
| Amiens | 1220 | 1269 | 49 |
| Sait-Jean Baptiste d'Aire | 1100 | 1400 | 300 |
| Agen | 1200 | 1900 | 700 |
| Ajaccio | 1577 | 1593 | 16 |
| Albi | 1282 | 1480 | 198 |
| Angers | 1200 | 1300 | 100 |
| Angouleme | 1100 | 1128 | 28 |
| Autun | 1120 | 1146 | 26 |
| Auxerre | 1215 | 1550 | 335 |
| Avignon | 1150 | 1425 | 275 |
| Bayeux | 1050 | 1450 | 400 |
| Bayonne | 1213 | 1615 | 402 |
| Bazas | 1200 | 1300 | 100 |
| Beauvais | 1225 | 1600 | 375 |

I looked into the distribution of how many years it took from the inception of the construction until the construction was declared completed. Figure 2 shows the results of the analysis with a median value of 263 years and mean value of 275 years. A group of outliers with construction durations between 600 and 700 years exists, notably the cathedrals in Agen, Cologne, and Rouen.

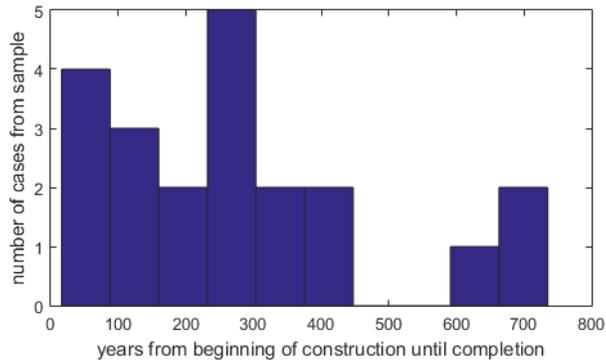

*Figure 2: French cathedrals and their years of construction*

The results confirm that cathedral-building was indeed an intergenerational endeavour and on average took 2 to 3 centuries to complete. This result is remarkably similar to results reported for a sample of English cathedrals in [3, p.39] of 250 to 300 years.

However, these results do not provide insights into whether the construction process was continuous or not. There is evidence that although construction was in many cases, continuous, intensity strongly varied. Often, construction went through periods of intense activity and long periods of moderate activity. Periods of intense activity were likely due to available funding [5, p.42] but also civil unrest, wars, plagues, [5, p.38]. Prak [23, p.387] notes that the construction of Canterbury Cathedral comprised 161 years of high activity and 182 years of low activity. Plotting the data from [5, p.40] for the Canterbury Cathedral's successive active construction periods and periods of inactivity results in Figure 3. The lengths of these periods do not seem to follow a recognizable pattern.

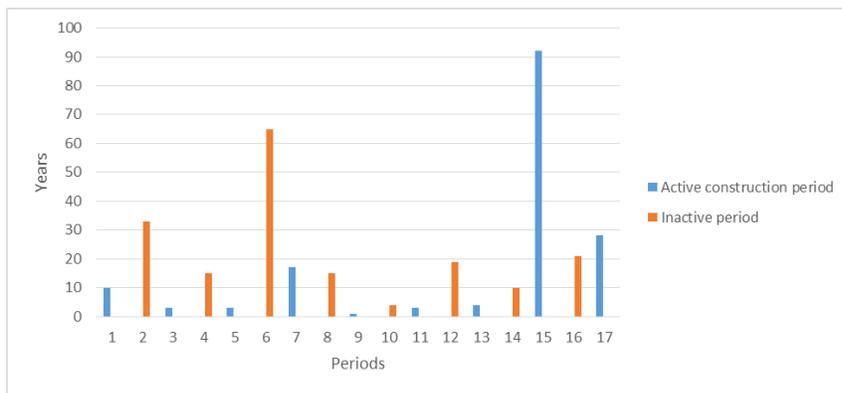

*Figure 3: Active and inactive periods of the construction of the Canterbury Cathedral*

Counting the occurrence of active periods from longer to shorter ones results in Figure 4.

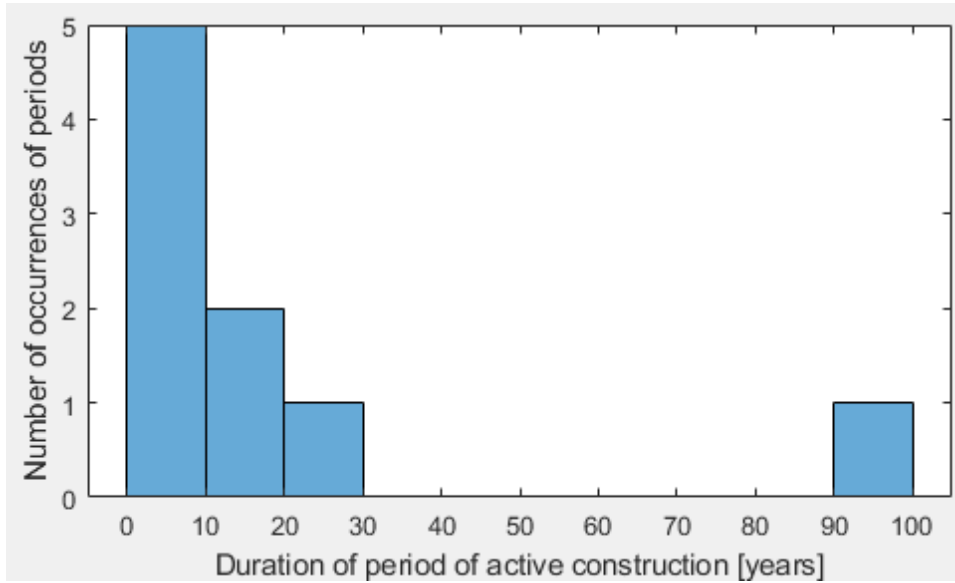

*Figure 4: Distribution of active periods in the construction of the Canterbury Cathedral*

Doing the same for inactive periods results in Figure 5.

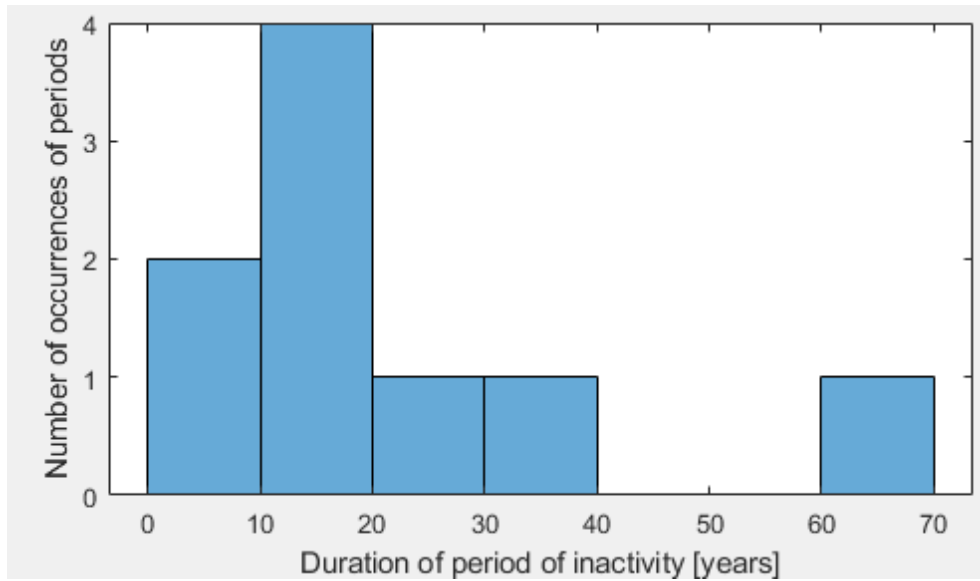

*Figure 5: Distribution of inactive periods in the construction of the Canterbury Cathedral*

Further data for Osnabrück and Utrecht Cathedral is taken from Vroom [13]. Periods in which funding is spent on construction and furnishing are counted. The results for all three data sets can be seen in Figure 6 for the active years and in Figure 7 for the inactive years for all three cathedrals. It can be seen that the length of the vast majority of construction periods falls between 1 and 30 years. Longer construction periods, however, exist (between 60 and 70 years and 90 to 100 years).

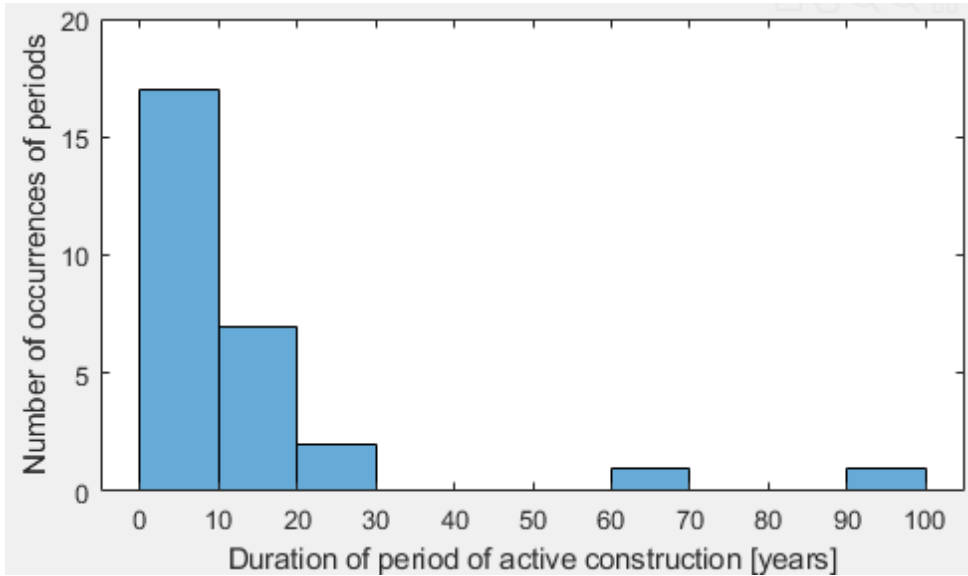

*Figure 6: Distribution of active constructions periods for Canterbury, Osnabrück, and Utrecht Cathedral*

A similar distribution is obtained for inactive periods, where the vast majority of periods falls between 1 and 40 years.

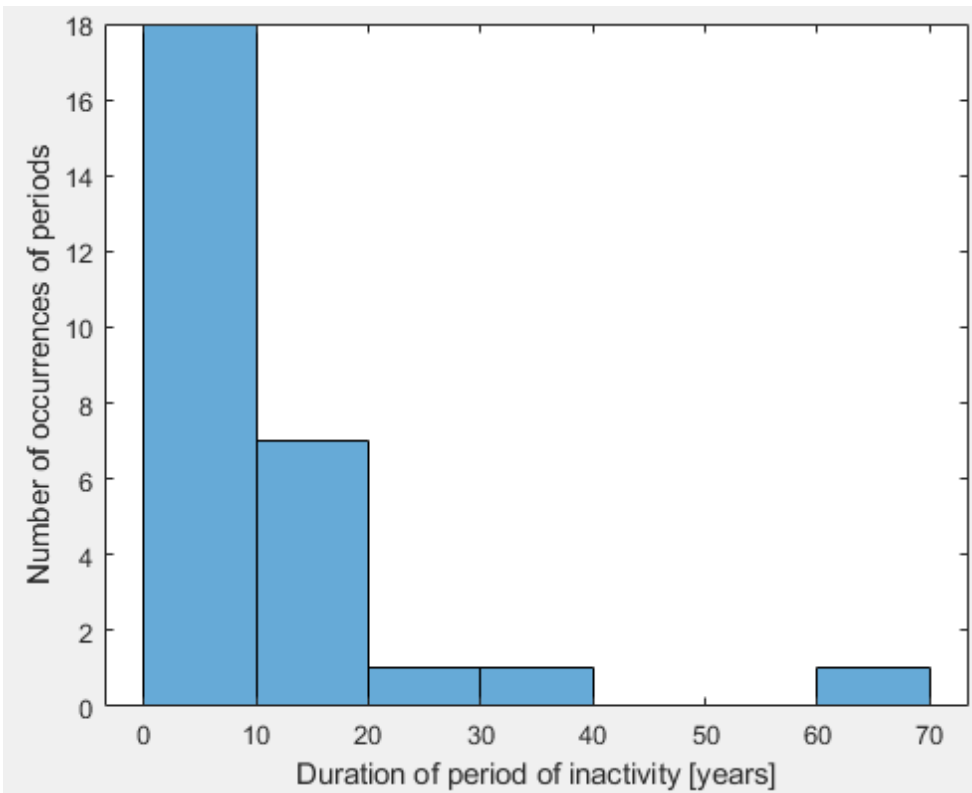

*Figure 7: Distribution of inactive periods for Canterbury, Osnabrück, and Utrecht Cathedral*

Fitting both distributions with a power law and exponential curve leads to an R² of 0.97 and 0.99 for active periods and similarly, 0.97 and 0.99 for periods of inactivity. Such distributions might be linked to the probabilistic distribution of weather phenomena such as drought. If there is a correlation between the two, this would confirm Scott's argument that funding uncertainties are linked to the occurrence of harmful weather conditions, which can be considered random [5].

To summarize, the results imply that active construction periods of cathedrals typically had a duration between 1 to 30 years with periods of inactivity of similar length between them. Only in rare cases are active and inactive periods of longer duration observed. This confirms that cathedrals were not built continuously until completion, except for rare cases such as Notre-Dame de Paris.

## 5. Were investments continuously provided?

A common conception within the space domain seems that cathedral building was financed by the public in the most general sense. Fortunately, there are several publications that took an in-depth look into the financing of cathedral-building in the Middle Ages. Kraus [14] analyzes the economics of cathedral-building and identifies factors that have contributed to a relatively short building duration. There are factors which pertain to specific stakeholders, such as a strong commitment by the ecclesial family, which was by far obvious. Some bishops had no interest in cathedral-building and did not commit personal funds to the undertaking. In the seminal book "Financing Cathedral Building in the Middle Ages" [13] Wim van Vroom identifies numerous funding sources for cathedrals that were built north of the Alps, notably the bishop, chapter, pope, king or lord, municipalities, and the faithful in general. Interestingly, funds from the bishop and chapter were only important when construction was initiated. On the long run, the contributions from the faithful of the diocese comprised the majority of the funds for construction. As the faithful of the diocese were a large and geographically distributed group, systematic and long-term fundraising campaigns were organized. The two main incentives provided were indulgences and relic worshipping. Indulgence trade became a huge economic activity, and relic worshipping was organized systematically to increase its efficiency. These two activities were accompanied by collecting campaigns within the diocese.

An interesting aspect is what fraction of total economic activity has been devoted to cathedral-building. Lopez [16] argues that the poor performance of urban economies north of the Alps compared to those in Italy was due to the large resource consumption of cathedrals. Vroom demonstrates that this thesis cannot be maintained and demonstrates that the average expenditures for the cathedral at Utrecht, amounted to the equivalent of the wages of 81 unskilled workers its construction from 1395 to 1527 [13]. Values of the same order of magnitude are displayed by Vroom for the Exeter Cathedral, Osnabrück Cathedral, Segovia, Sens, and Troyes Cathedral. These yearly expenses were far below those of palaces, castles, and fortifications. Hence, cathedral-building seems to be far less resource-consuming than one might expect.

Regarding the financial stability of cathedral-building, Vroom concludes that "Observed over the long term, the general pattern of cathedral fabric expenditures paints a picture of instability" [12, p.462] "In Durham, one chronicler tells us, the pace of construction in the early twelfth century rose and fell with the magnitude of the offerings made there."[12, p.115] Scott [5] ties instabilities in cathedral funding ultimately to the instability of agricultural yield in the Middle Ages, the main source of income. Weather anomalies such as droughts could severely impact income, which then reduced the amount of funding available for cathedrals.

However, over time, mechanisms were put in place to stabilize cathedral financing. During the thirteenth century, a church office, the vestry, was created for managing the fabric. The vestry was administered by the cathedral chapter. Bishops and canons were then "obliged to give it a fixed proportion of the church income" providing at least some stability to financing cathedral construction [6, p.272]. Furthermore, "Sometimes, these miracles were connected with the building works." [12, p.116]

Hence, financing for cathedrals was far from stable, although mechanisms were developed over time to mitigate these risks, notably via establishing the vestry, which actively managed the cathedral fabric.

## 6. Discussion

The previously presented results confirm that cathedral building cannot serve as an example for a successful, continuous, large-duration project. The results seem to imply something far more interesting. Cathedral building was confronted with an "extreme complexity of the project, the lengthy period of time required for construction, and the repeated interruption of the building process" [5, p.140]. These are characteristics, which are often ascribed to space programs [18]. In the following, I will discuss what mitigation strategies cathedral builders used and how these strategies might be transferred to the context of space programs.

**Extensibility of the cathedral architecture**

As Lopez [15, p.273] describes, cathedrals needed to provide the possibility for church service throughout their whole construction period. "Cathedrals-in-progress were given temporary wooden roofs, and makeshift services were conducted in these half-built structures." "These services were separated from the building work by temporary screens, some of them rather robust structures that were never subsequently removed." The ability of the cathedral to enable church service throughout its construction was probably crucial in sustaining construction work over decades and centuries. At any point, the church could be used for its intended purpose. In today's marketing terms, one can talk about a minimum-viable product (MVP) (cathedral that allows for church service) to which new features (new elements of the cathedral) are added. There are elements of the cathedral that are obviously modular such as the tower(s), the spire, and chapels. These elements can be added without interfering with church services. Fig. 1 shows the floorplan of Chartres Cathedral, where the side chapels can be seen as the half-circles and rectangles on the top and on the left.

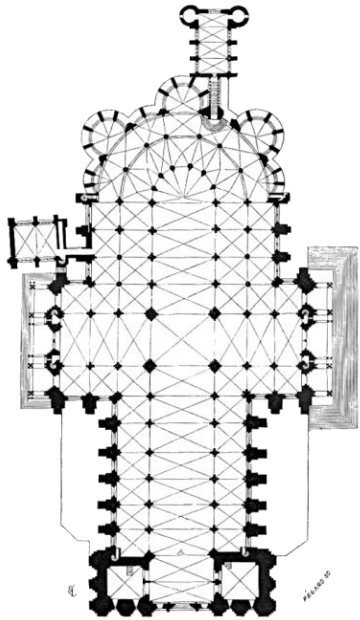

*Fig. 1: Floorplan of the Chartres Cathedral (Wikipedia)*

Scott [5, p.142] even goes further than that and argues that modularity was a key enabler for cathedral building over its construction duration of decades to centuries. Modularity was not only present in the large elements (tower, nave, etc.) but also in sub-elements such as columns, arcs, etc. This allowed for a sub-division of work down to individual stones, robust against various uncertainties and interruptions in construction. Interestingly Scott takes inspiration from Herbert Simon regarding modularity and how it is able to mitigate uncertainties.

**Trans-generational transfer of technological capabilities**

According to Billington and Mark [19], "the great cathedrals were constructed by relatively well-paid, highly skilled teams of masons and carpenters, with supporting staffs that included the apprentices who insured the continuity of skills" [6, p.38] According to White [20], "we see a sublime fusion of high spirituality and advanced technology." Indeed, it seems that the investments from the faithful were channelled into one of the few areas during the Middle Ages where considerable technological progress was made by constantly developing and transferring skills in diverse art forms but also civil engineering and architecture [21]. One notable document in this respect is the only written first-hand account of the engineering and architecture behind Gothic cathedrals, is the Carnet of Villard de Honnecourt [22]. Furthermore, just to provide one example for the engineering ingenuity of the Middle Ages: The flying buttresses were introduced as structural support for very tall vaulted structures against horizontal loads such as winds [19,23]. Little seems to be known about the process of how knowledge was transferred between generations of builders [5,11]. Turnbull [11] argues that no global and accurate plan existed, but low-level templates were used. Scott [5] discusses the possibility that modularity allowed for continuity in building activity in the absence of a master plan. Hence, it can be speculated that cathedral building was robust against knowledge loss by providing a modular architecture where new contributions could be "plugged in".

## 7. Cathedrals and starships: Lessons learned?

I will now suggest some conclusions from the cathedrals to long-term, large-scale space projects, exemplified by the starship.

From the previous sections we can summarize the following characteristics of cathedral-building as a long-term, large-scale engineering project:

- *Value-delivery via a minimum viable product:* The way cathedrals were built ensured that the main function of the cathedral, church service could be performed throughout the whole construction period. Hence, continuous value delivery was ensured.
- *Project finance:* Cathedral-building was subject to instabilities in available funding. However, organizational innovation mitigated this risk. Formal institutions and mechanisms were created to increase the reliability of cash flows. Fundraising campaigns along with spiritual "products" such as indulgences, relics, and miracles were vital elements of the financing strategy.
- *Extensibility:* The architecture, in the sense of the components and relationships between these, enabled the continuous addition of new components (towers, chapels, flying buttresses) to extend the church. Hence, the difficulty in determining if a cathedral construction is ever "finished".
- *Knowledge transfer:* Knowledge transfer by or between key personnel such as architects and artisans, was facilitated by the dissemination and documentation of their knowledge. Furthermore, this workforce was highly mobile, enabling the cross-fertilization of building projects in different regions and even countries. These enabled not only the transfer of explicit knowledge in the form of documents but also the transfer of tacit knowledge.

- **Stakeholder management:** The network of stakeholders around a cathedral construction project contained different entities such as the ecclesiarchy, municipalities, representatives of worldly powers, the construction workforce, and the faithful in the diocese. The management of the diverse value flows, and notably, financial flows seem to have been as complex as the physical cathedral.
- **Expenditures:** The financial and human resources spent on cathedrals were much smaller than for palaces, castles, and fortifications. The expenditures were typically on the order of 10s to 100s of annual wages with large gaps between periods of activity [13].

From these characteristics, I can now speculate on lessons to be learned for future long-term, large-scale space projects. Rather than suggesting specific solutions, I will discuss general principles and how they might translate into solutions for such projects:

- **Modular architecture:** Modular architectures do not only exist for buildings. Modularity also exists for knowledge, software, etc. Cathedrals somehow provided interfaces where future builders could continue their work, such as wall and roof sections. Hence, on the one hand, they could build on previous work or, on the other, they could do so efficiently via building on these interfaces. Furthermore, this approach left important degrees of freedom, so that future builders had relative freedom in their designs. Today, we would call such an architecture an *open* architecture. Modules can be integrated via well-defined interfaces. An open architecture approach ensures flexibility and extensibility, able to absorb uncertainties. In the context of starship development, subsystem technologies could be matured independently but with a clear definition of interfaces and constraints. Such an approach would ensure that these technologies can be integrated at some point in the future. "Interfaces" in this case should be understood broadly, such as the fundamental compatibility of technologies, information exchange between organizations which develop the technologies, etc.
- **Value delivery / minimum viable product:** Even for long-term projects, thinking about how intermediate value can be delivered is important. For example, for interstellar travel, an intermediate version of the propulsion system could be used for interplanetary travel, as has been suggested for laser sail propulsion [24–26].
- **Financial stability:** Funding for a starship project is likely to be unstable, which has been demonstrated repeatedly for past and current space programs. What institutional and financial mechanisms could be developed to mitigate the risk from instability? The example of the cathedral has shown that financial stability was established over time via institutions. Recently, foundations and non-profits have emerged which provide funding for research on interstellar travel such as the Breakthrough Foundation and the Limitless Space Institute. Compared to the Middle Ages, a vast number of mechanisms and instruments for financing long-term projects exists today.

## References

[1]     M. Schiller, Benefits and Motivation of Spaceflight, (2008). http://mediatum2.ub.tum.de/?id=654918 (accessed December 31, 2016).

[2]     M. Griffin, Space exploration: real reasons and acceptable reasons, Quasar Award Dinner, Bay